# A Machine Learning Framework for EEG-Based Prediction of Treatment Efficacy in Chronic Neck Pain

Xiru Wang[1,*], Aiden Li[2], Hongzhao Tan[2], Stevie Foglia[1], Aimee Nelson[1,*], and Zhen Gao[2,*]

[1]*Department of Kinesiology, Faculty of Science, McMaster University, Hamilton, ON, Canada*
[2]*W Booth School of Engineering Practice and Technology, Faculty of Engineering, McMaster University, Hamilton, ON, Canada*

***Correspondence:** wangx256@mcmaster.ca; nelsonaj@mcmaster.ca; gaozhen@mcmaster.ca

## Abstract

Chronic neck pain is a leading cause of disability worldwide, and current treatment selection remains largely trial-and-error. We present a machine learning framework that uses electroencephalography (EEG) to predict treatment efficacy in patients with chronic neck pain, with the goal of supporting individualized therapy and reducing the burden on healthcare systems. The framework centres on a rigorous data preprocessing stage tailored to the characteristics of each EEG recording type. For resting-state EEG, the preprocessing pipeline comprises baseline signal removal, bad channel identification and exclusion, re-referencing, bandpass and notch filtering, Independent Component Analysis (ICA), and power spectral density (PSD) analysis. For motor execution and motor imagery recordings, the same initial steps are applied, after which signals are aligned to trigger events (typically from 0.5 s before to 1 s after each trigger) so that event-related desynchronization (ERD) and event-related synchronization (ERS) can be quantified. Synchronously recorded electromyography (EMG) data are bandpass-filtered and smoothed with a moving average, then correlated with the corresponding EEG channels to characterise the EEG–EMG relationship during attempted movement. In parallel, we performed an extensive literature review of machine learning models applied to clinical EEG (763 records initially screened, 16 patient and 47 healthy-control studies retained), to inform the post-processing strategy. Through this combined preprocessing and review effort, we aim to develop a robust predictive model that can support personalised healthcare strategies in chronic pain management.

**Keywords:** Electroencephalography (EEG); chronic neck pain; data preprocessing; Independent Component Analysis; event-related desynchronization; machine learning; multimodal data fusion.

## 1. Introduction

Chronic neck pain is a prevalent and debilitating musculoskeletal condition, ranking among the leading global causes of years lived with disability [1, 2]. Evaluating treatment efficacy in patients with chronic neck pain is complex: response to physical therapy, pharmacotherapy or interventional procedures is heterogeneous, and patients often cycle through multiple options before finding one that provides meaningful relief [3]. This trial-and-error pathway prolongs suffering and places a substantial cost burden on the Canadian healthcare system [3, 4]. The present study aims to optimise resource allocation by accurately predicting treatment effectiveness for individuals with chronic neck pain from non-invasive neurophysiological recordings.

Electroencephalography (EEG) is an attractive modality for this task: it is inexpensive, portable, has millisecond temporal resolution, and provides direct access to the cortical dynamics that are altered in chronic pain states [4, 5]. A growing body of work has identified candidate EEG biomarkers of chronic pain, including changes in resting-state band power, peak alpha frequency, and inter-regional connectivity [4, 5, 6]. In parallel, deep learning has matured into a practical tool for EEG analysis, with convolutional, recurrent, and transformer architectures all demonstrating competitive performance on EEG classification and decoding problems [7].

By applying machine learning, and in particular deep learning, to multi-channel EEG recorded from patients with chronic neck pain, we seek to predict treatment outcomes and offer a quantitative complement to clinical decision making. A fundamental component of this work involves preprocessing EEG data to ensure its suitability for model training. ICA plays a critical role in this process [8, 9], enabling the removal of artifacts and yielding a clean dataset for subsequent model development. In addition, a comprehensive understanding of deep learning applications in EEG signal analysis is essential. To this end, we conduct an extensive literature review to identify best practices and derive insights that inform our model development and validation strategies [7].

The multivariate, high-dimensional, and noisy nature of multi-channel EEG introduces additional complexity, presenting challenges for both model construction and training. Nevertheless, the potential to accurately predict treatment efficacy for patients with chronic neck pain highlights the significance and urgency of this research. Through interdisciplinary collaboration and rigorous methodological approaches, we aim to contribute to improved healthcare outcomes and more efficient resource utilisation in chronic pain management.

## 2. Data Preprocessing Techniques

EEG signals were recorded from 64 scalp electrodes, with each electrode placed at a defined position on the participant's scalp according to the international 10–20 system [10]. The recording protocol was divided into four segments: eyes open / eyes closed (resting-state), hand clench (reactivity), neck movement, and imagined hand clench (motor imagery). For preprocessing, these four conditions were grouped into two categories: (i) resting-state data, comprising the eye open/close recordings, and (ii) motor execution / motor imagery data, comprising hand clench, imagined hand clench, and neck movement recordings.

The common preprocessing steps applied to both categories, in order, were baseline signal removal, bad channel removal, re-referencing, bandpass and notch filtering, and Independent Component Analysis (ICA) [8, 9]. After these shared steps, power spectral density (PSD) analysis was computed for the resting-state data, while event-related desynchronization (ERD) and event-related synchronization (ERS) were computed for the motor execution / motor imagery data [11, 12].

### 2.1 Baseline Signal and Bad Channel Removal

Both baseline signal removal and bad channel removal were performed to remove segments of the raw EEG that were corrupted at the start of recording or by faulty electrode contacts. During initial visualisation of the raw EEG, the absolute amplitudes for the first 60 s (or longer) of every recording were two to three orders of magnitude larger than those of the remaining recording across all channels. This is a known consequence of the recording software's initial baseline-finding routine, during which it requires several seconds to converge on a stable reference for the recorded electrical signals. Baseline signal removal was therefore applied to truncate the first 60 s (or longer, where required) of each raw EEG recording.

During recording, the contact between certain electrodes and the scalp can be physically unstable, particularly for participants with thick hair, where electrodes placed at the back of the scalp tend to make poor contact. Such low-quality channels typically produce signals one to two orders of magnitude larger in amplitude than well-connected channels, contain little or no neural information, and would distort downstream preprocessing if retained. Data from these "bad channels" were therefore removed before further preprocessing.

To statistically identify bad channels from raw EEG data, the standard deviation of the EEG data from each of the 64 channels was computed; channels whose standard deviation was at least twice that of the remaining channels were flagged for removal. As an additional rule of thumb, scalp EEG amplitudes that consistently exceed 200 μV were also treated as artifactual [10].

## 2.2 Re-referencing

EEG values are, by definition, differences in electric potential (in microvolts) between one or more reference channels and each of the remaining channels: the data from the reference channel(s) are subtracted from the data of every other channel. The choice of reference channel(s) is therefore consequential, since an ideal reference captures little brain-specific activity and primarily reflects environmental noise that is also present in the recording electrodes. Among the 64 channels, two channels A1 and A2 placed on the earlobes (Fig. 1) were used as reference channels, in line with common practice for clinical and research EEG recordings [10].

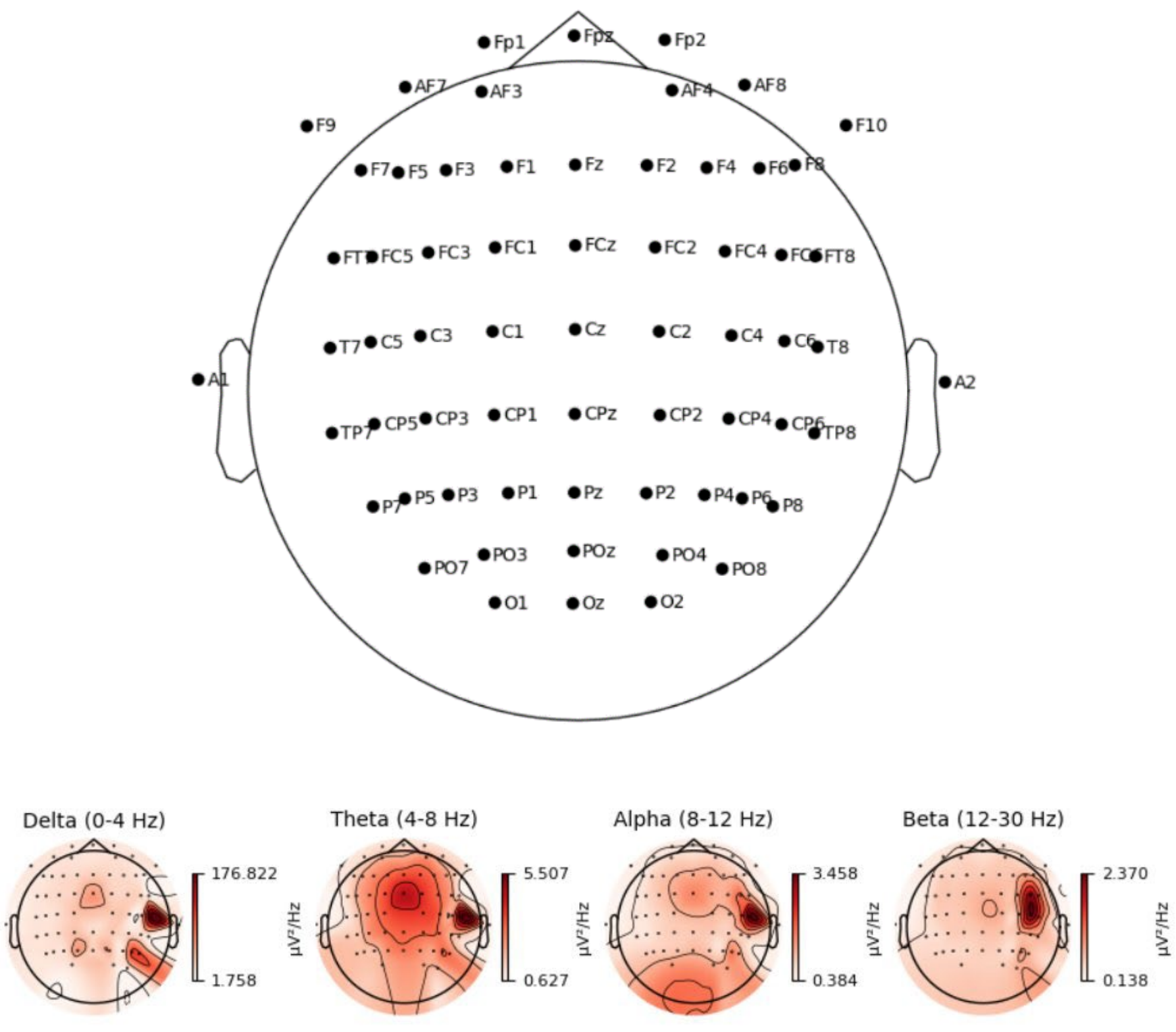

**Figure 1.** Standard 10–20 system montage for the 64-channel recording configuration used in this study. The two earlobe electrodes A1 and A2 (visible at the left and right of the head outline) serve as references for the remaining channels.

## 2.3 Bandpass and Notch Filters

After re-referencing, bandpass and notch filtering were applied. EEG is a complex electrical signal that can be represented as a sum of sinusoids at different frequencies. Scalp EEG of physiological interest is mostly contained within the 0.1–30 Hz band [10]. Recordings, however, also contain non-neural fluctuations: low-frequency artifacts originating from head and electrode-wire movement, and higher-frequency artifacts originating from sources such as facial muscle contractions. To attenuate these out-of-band components, a bandpass filter was applied with a lower passband edge of 0.1 Hz and an upper passband edge of 30 Hz.

In addition to broadband bandpass filtering, EEG recordings often contain narrowband line-frequency interference from power lines and nearby electrical equipment. In North America this interference is centred at 60 Hz, so a notch filter at 60 Hz was applied to suppress it without affecting the broader spectrum of interest.

## 2.4 Independent Component Analysis

During EEG recording it is virtually impossible to avoid artifacts, defined as recorded activity that does not originate from the cerebrum. Artifacts may be physiological in origin, such as eye blinks, eye movements, or swallowing, or non-physiological, such as movements of the electrodes relative to the scalp [9, 13]. Independent Component Analysis (ICA) is an algorithm for estimating statistically independent source signals that have been mixed in unknown proportions, and is widely used in EEG preprocessing to isolate and remove artifactual sources [8, 9].

Prior to applying ICA, the EEG data were scaled to unit variance and decomposed using Principal Component Analysis (PCA). The first N principal components capturing the highest fractions of variance were then passed to ICA to obtain Independent Components (ICs). Artifactual activity is typically concentrated in a subset of these ICs. Each component was therefore assessed, either using automated approaches or by manual visual inspection, and components carrying artifacts (e.g., component ICA000 in Fig. 2, which exhibits the high-amplitude transients characteristic of eye-blink and movement artifacts) were excluded. The cleaned EEG was then reconstructed from the retained ICs together with any principal components that had not been passed into ICA.

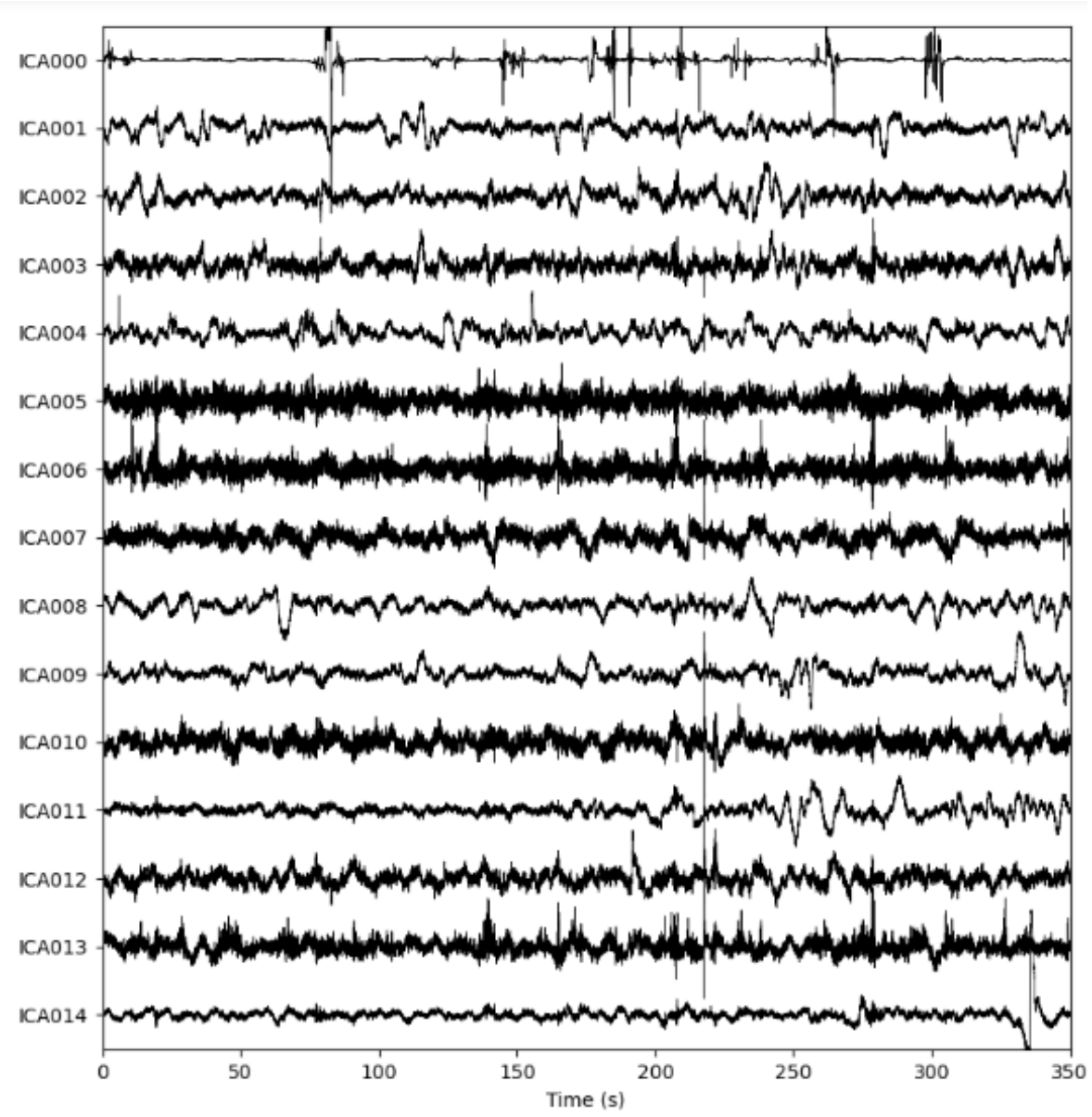

**Figure 2.** Time-courses of 15 Independent Components (ICA000–ICA014) decomposed from a single subject's EEG recording. Component ICA000 (top trace) exhibits high-amplitude transient deflections characteristic of eye-blink and movement artifacts, and would be excluded from the reconstructed signal.

## 2.5 Power Spectral Density Analysis

After artifact repair, resting-state EEG was characterised in the frequency domain over four classical bands: Delta (0–4 Hz), Theta (4–8 Hz), Alpha (8–12 Hz), and Beta (12–30 Hz) [10]. The Fast Fourier Transform (FFT) was used to map the time-domain signal into the frequency domain, and the squared magnitude of the FFT yielded an estimate of the power spectral density (the periodogram) in units of μV²/Hz. Welch's method [14] was used in place of the classical periodogram, since segment-averaged modified periodograms reduce variance at the cost of a modest reduction in frequency resolution. The per-channel powers were averaged within each of the four frequency bands to obtain mean band powers.

These mean band powers serve directly as input features for the downstream deep-learning model, and can additionally be visualised as scalp topographies (one heatmap per frequency band) on the standard 10–20 montage (Fig. 3), since the 64 recorded channels form a superset of the 10–20 layout [10].

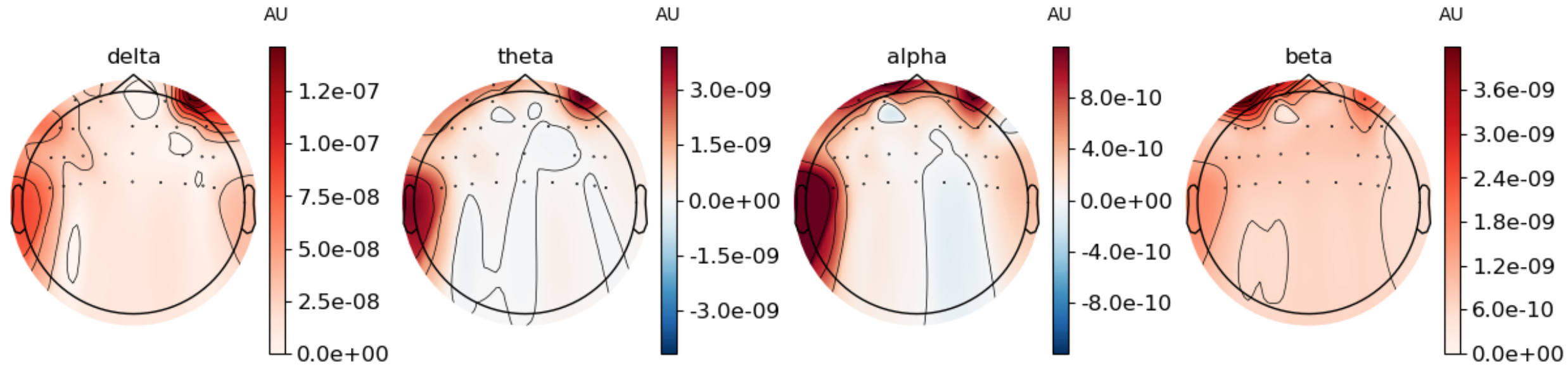

**Figure 3.** Scalp topographies of average band power computed on resting-state EEG for the four classical frequency bands (Delta 0–4 Hz, Theta 4–8 Hz, Alpha 8–12 Hz, Beta 12–30 Hz). Colour scales are in μV²/Hz.

A complete preprocessing pipeline for the resting-state EEG was implemented in Python within the Jupyter Notebook environment. EEG recordings were imported from HDF5 files using the h5py library. Most preprocessing operations relied on the MNE-Python package [15], an open-source toolbox for the analysis of M/EEG and other neurophysiological signals. Pandas, Matplotlib, and Plotly were used for visualisation of raw and processed data after each preprocessing step. Specifically, finite-impulse-response (FIR) filters were used for both bandpass and notch filtering, and Welch's method [14] was used for power spectral density estimation. Because electrooculography (EOG) channels were not available in the present recordings, the team manually inspected the ICs produced by ICA to identify and reject artifactual components, rather than relying on EOG-based automatic classifiers.

### 2.6 Event-Related Potentials and ERD/ERS

Event-related potentials (ERPs) are neurophysiological signals extracted from the EEG that reflect the brain's electrical response to specific sensory, cognitive, or motor events [16]. Unlike the ongoing EEG, which captures continuous brain activity, ERPs are stimulus-specific waveforms time-locked to events, and are particularly useful for characterising how the brain processes different stimuli over time. To obtain an ERP, a participant is repeatedly exposed to a stimulus while EEG is recorded; the ERP is then extracted from the ongoing EEG by aligning and averaging segments time-locked to the stimulus. Averaging enhances activity related to the event and attenuates uncorrelated background noise, which averages towards zero. ERPs are characterised by their polarity (positive or negative), amplitude, and latency (the time relative to stimulus onset). In the context of chronic pain, ERP studies have revealed alterations in the way pain-related signals are processed in the brain — for example, certain ERP components are delayed or attenuated in chronic pain populations, suggesting altered neural processing pathways [4, 5].

To extract ERP-related features from our EEG data, we implemented a dedicated preprocessing pipeline for the motor imagery / motor execution recordings. During these recordings, a trigger marker indicates the moment a stimulus was presented, and the participant either performs or imagines the corresponding action (e.g., turning the head to the left). The pipeline performs two complementary analyses:

- Computation of changes in EEG band-power that are time-locked to the trigger. Decreases in power are termed event-related desynchronization (ERD), and increases are termed event-related synchronization (ERS) [11, 12].
- Computation of correlations between the alpha-band EEG signal of each channel and the synchronously recorded electromyography (EMG) signal, following the pipeline of Chowdhury et al. [17].

To quantify event-related power changes, the bandpass-filtered, notch-filtered, ICA-cleaned EEG was first segmented into epochs around each trigger (e.g., from 1 s before to 2 s after each trigger). For every epoch, every channel, and every 1 Hz frequency bin from 1–30 Hz, a spectrotemporal (time–frequency) representation was computed, using the average band-power within the −1 to 0 s pre-stimulus window of each epoch as the baseline for baseline correction [11, 12]. Line plots and scalp heatmaps were then produced by averaging the spectrotemporal representations across epochs and across the bins within each of the four canonical bands (Delta, Theta, Alpha, Beta) (Figs. 4 and 5).

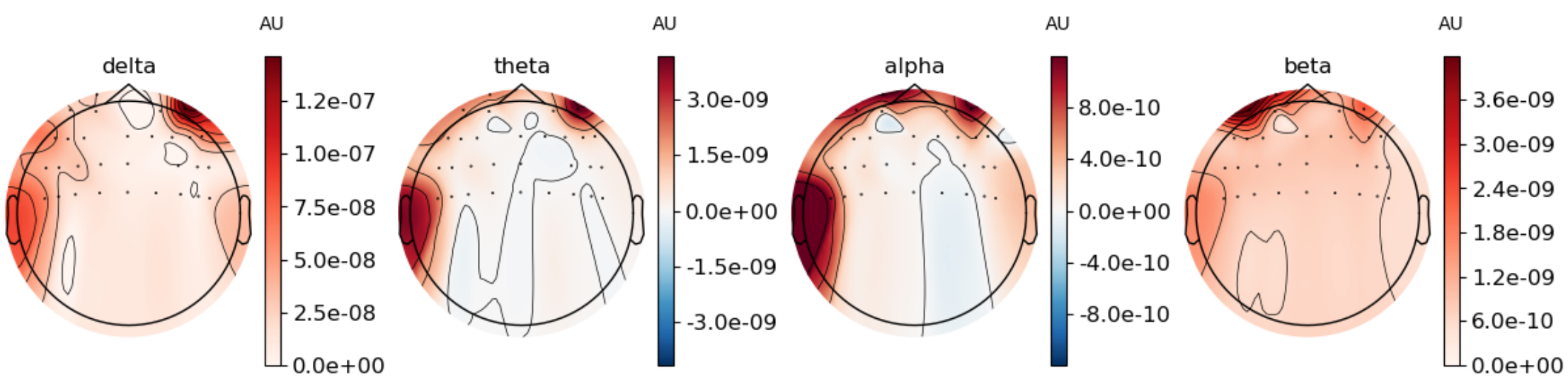

**Figure 4.** Scalp topographies of trigger-locked spectrotemporal power for the four frequency bands (delta, theta, alpha, beta). Topographies are computed by averaging baseline-corrected spectrotemporal representations across epochs and across all 1 Hz bins within each band. Colour scale in arbitrary units.

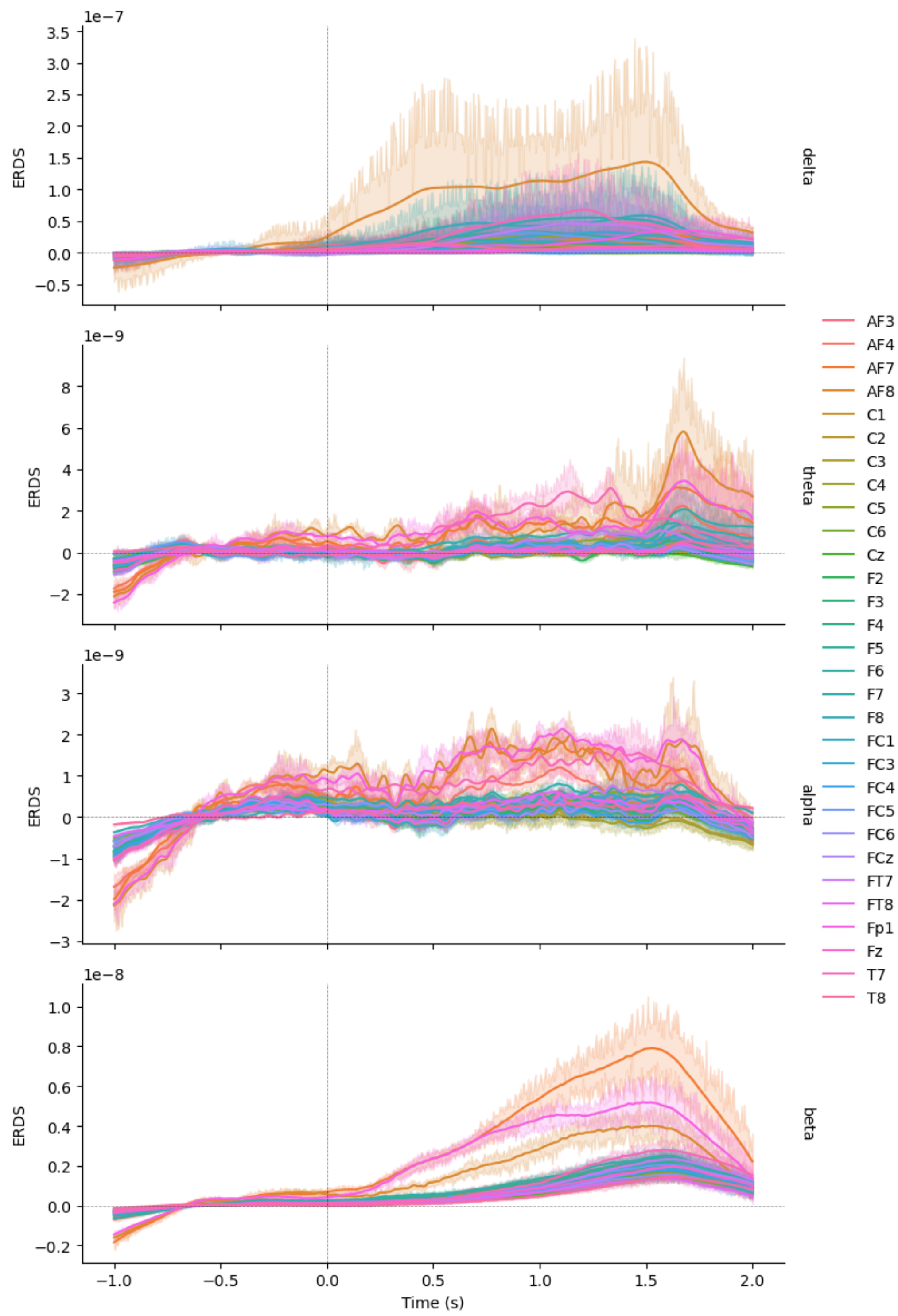

**Figure 5.** Per-channel time-courses of trigger-locked event-related desynchronization/synchronization (ERDS) for delta, theta, alpha and beta bands. Time 0 marks trigger onset; the −1 to 0 s window is used as the baseline. Each coloured line corresponds to one EEG channel.

EMG measures the electrical activity generated when motor neurons stimulate skeletal muscle fibres. Higher absolute correlations between the EEG of a given channel (or group of channels) and the EMG signal during a trigger-locked movement attempt imply tighter cortico-muscular coupling, and therefore better cortical control of the muscle in response to the cue. Following baseline-signal and bad-channel removal, we implemented additional EEG preprocessing steps and a parallel EMG preprocessing pipeline modelled on the procedure of Chowdhury et al. [17]. Within each epoch we extracted the preprocessed EEG and EMG samples falling inside a short window (e.g., [−0.5, 0.5] s, where the end of the interval was constrained to remain within the epoch) around the time at which the EMG power first exceeded a threshold (e.g., 500 $\mu V^2$); these samples were then used to compute the correlation between EMG and each EEG channel (Fig. 6). Heatmaps of the absolute correlation values, averaged across all epochs, were produced to visualise the spatial pattern of cortico-muscular coupling across the scalp (Fig. 7).

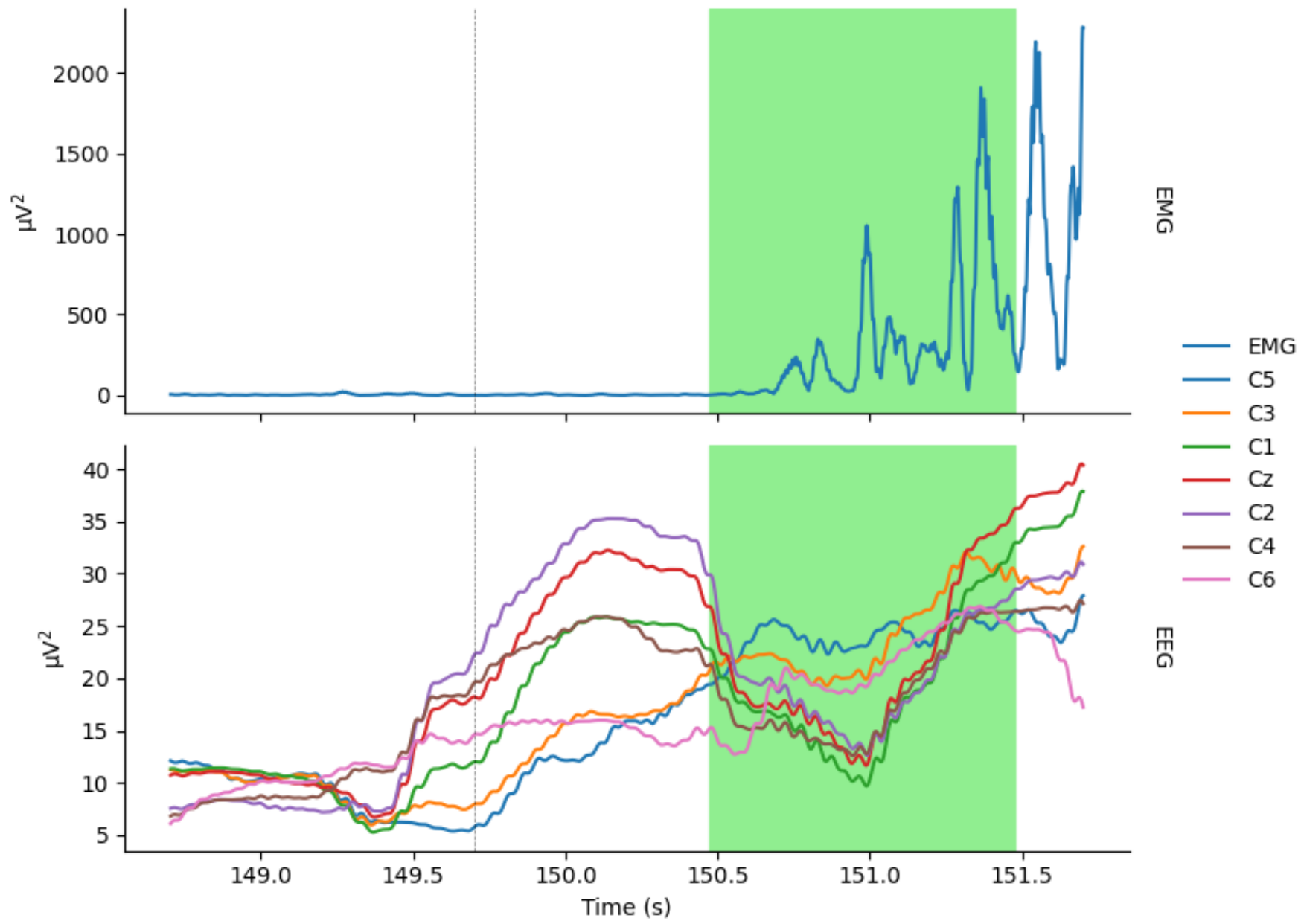

**Figure 6.** Preprocessed EEG and EMG signals within a single trigger-locked epoch. The dashed grey vertical line marks trigger onset. The green-shaded region indicates the analysis window over which EEG–EMG correlations are computed. Top: EMG channel; Bottom: a subset of central EEG channels (C1–C6, Cz).

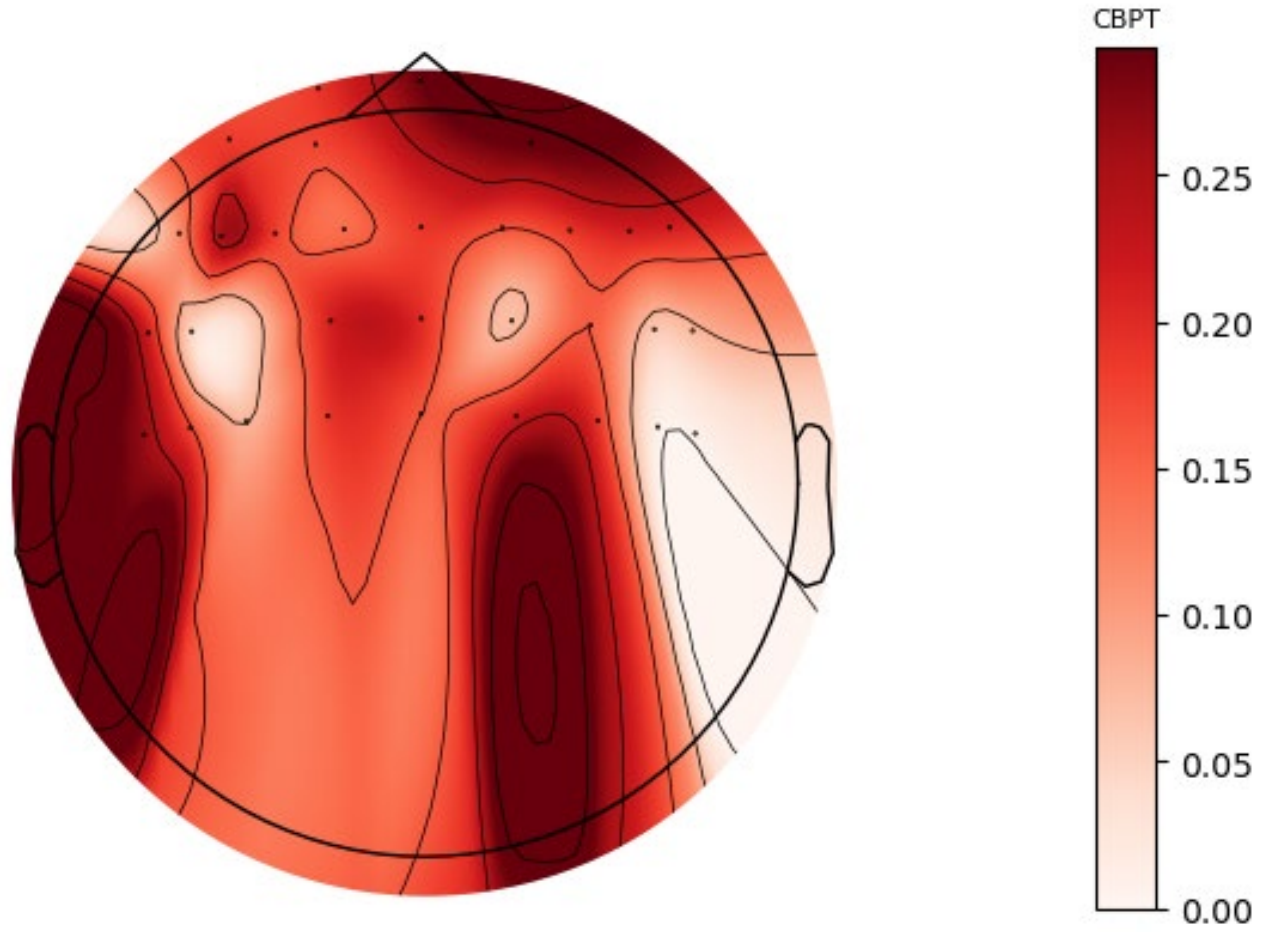

**Figure 7.** Scalp topography of absolute EEG–EMG correlation values, averaged across all epochs, for each of the recorded EEG channels. Higher values (darker red) indicate stronger cortico-muscular coupling during the cued movement attempt.

## 3. Literature Review for Data Postprocessing

To inform the choice of machine learning models for the post-processing stage, we conducted a structured review of recent EEG and multimodal-EEG studies. The review proceeded by defining core topics, target clinical populations, candidate computational models, and acceptable data-management strategies. An initial pool of 763 papers was assembled from PubMed, Web of Science, and Google Scholar. After title screening, 184 papers remained; after abstract screening, 103; after full-text screening, 81. Following team discussion, we retained 16 papers reporting on patient cohorts and 47 papers reporting on healthy-control cohorts as the working corpus for downstream analysis. The full filtering workflow is summarised in Section 3.1.

### 3.1 Clinical Tasks and Their Relevance

The retained papers covered a broad spectrum of clinical tasks, including autism spectrum disorder (ASD) detection, seizure detection, epilepsy monitoring, mild depression analysis, and anxiety assessment, among others. Despite this clinical heterogeneity, the unifying methodological feature is the use of EEG (alone or in combination with other signals) to enhance prediction accuracy. Among these studies, those targeting pain-related conditions and motor imagery tasks are most directly relevant to our protocol, given the structure of our data-collection segments (hand clench, neck movement, imagined hand clench).

### 3.2 Machine Learning Models Overview

The machine-learning models employed in the reviewed papers vary substantially, reflecting the diversity of tasks and the heterogeneity of EEG data. A summary of model categories and their frequencies of occurrence in the corpus is given in Table 1.

| Category | Sub-Category | Machine Learning Models | Count | Sub-Count |
|---|---|---|---|---|
| Supervised Learning Models | Support-Vector Machine (SVM) | Linear Support-Vector Machine (SVM) | 7 | 12 |
| | | Multiple-kernel Support Vector Machine (MK-SVM) | 2 | |
| | | Radial Basis Function Kernel SVM (RBF-SVM) | 3 | |
| | Decision Trees | Decision Tree (DT) | 3 | 13 |
| | | Random Forest (RF) | 6 | |
| | | XGBoost | 2 | |
| | | AdaBoost (ADB) | 1 | |
| | | Gradient Boosting Decision Tree (GBDT) | 1 | |
| | Naïve Bayes | Naïve Bayes (NB) | 1 | 2 |
| | | Gaussian Naïve Bayes (GNB) | 1 | |
| | Discriminant Analysis | Logistic Regression (LR) | 1 | 2 |
| | | Quadratic Discriminant Analysis (QDA) | 1 | |
| | K-Nearest Neighbors | K-Nearest Neighbor (KNN) | 3 | 3 |
| Neural Networks | Convolutional Neural Networks (CNN) | Convolutional Neural Network (CNN) | 4 | 5 |
| | | ResNet | 1 | |
| | Recurrent Neural Networks (RNN) | Long Short-term Memory (LSTM) | 3 | 3 |
| | Multilayer Perceptron (MLP) | Multilayer Perceptron (MLP) | 1 | 3 |
| | | Self Normalizing Neural Networks (SNN) | 1 | |
| | | Batch Normalized Multilayer Perceptron (BNMLP) | 1 | |
| Advanced Models and Techniques | Transformers | Cyclic Transformer (CT) | 1 | 2 |
| | | Multimodal Input CT (MICT) | 1 | |
| | Hybrid and Specialized Models | ChronoNet | 1 | 3 |
| | | Elastic Net | 1 | |
| | | Penalized Logistic Regression | 1 | |

**Table 1.** Frequency of machine learning models used in the reviewed clinical EEG studies, grouped by category (Supervised Learning Models, Neural Networks, Advanced Models and Techniques) and sub-category.

### 3.3 Discussion of Commonly Used Models

From Table 1, decision-tree–based models (13 occurrences) and support-vector machines (12 occurrences) are the most frequently used algorithms in the reviewed studies. Their prevalence reflects their effectiveness on the high-dimensional, modestly-sized datasets that are typical of clinical EEG. We highlight several model families below.

#### *3.3.1 Support-Vector Machines (SVMs)*

Support-vector machines [18] are widely used for classification in high-dimensional spaces. Variants encountered in the corpus include linear, polynomial, radial-basis-function (RBF), and multiple-kernel SVMs, each suited to different data distributions and decision-boundary shapes. SVMs are robust to overfitting in high-dimensional, small-to-medium-sized datasets, which makes them a natural fit for typical EEG feature spaces.

#### *3.3.2 Decision Trees and Tree Ensembles*

Decision trees, and in particular random forests [19] and gradient-boosted variants such as XGBoost [20], are commonly applied to EEG classification and regression. Tree ensembles handle large heterogeneous feature sets, provide interpretable feature-importance rankings, and are robust to overfitting through bagging or boosting.

#### *3.3.3 Convolutional Neural Networks (CNNs)*

Convolutional neural networks appeared in 5 of the reviewed studies. CNNs are well suited to data with spatial or temporal hierarchies, including image data and time-series, and have repeatedly been shown to learn discriminative features directly from minimally pre-processed EEG, often outperforming hand-crafted features [7].

#### *3.3.4 Recurrent Neural Networks and LSTMs*

Recurrent neural networks, including Long Short-Term Memory (LSTM) networks [21], appeared 3 times. These architectures are designed for sequential and time-series data and excel at capturing long-range temporal dependencies, which are important when modelling the temporal structure of EEG signals.

#### *3.3.5 Other Notable Models*

Additional models in the corpus included K-Nearest Neighbours (KNN, 3 uses), the Multilayer Perceptron (MLP, 3 uses), and Transformer-based architectures including the Cyclic Transformer and the Multimodal-Input Cyclic Transformer (2 uses) [22]. Specialised hybrid models such as Self-Normalising Neural Networks (SNN), Batch-Normalised MLPs (BNMLP), ChronoNet, Elastic Net, and Penalised Logistic Regression each appeared once and address specific aspects of EEG data analysis (regularisation, normalisation, or interpretability).

### 3.4 Multimodal Data Integration

Combining EEG with other modalities can substantially improve the accuracy and robustness of predictive models in complex biomedical contexts such as chronic pain assessment [4, 5, 23]. The reviewed papers illustrate a wide variety of complementary modalities, summarised below, together with the integration (data-fusion) strategies that connect them to EEG.

#### *3.4.1 Multimodal Data Types*

- Facial expression and eye-fixation features. Behavioural indicators that contextualise EEG signals, particularly with respect to emotional state and cognitive load, and that may help disambiguate different pain states or stimulus responses.
- Functional and structural Magnetic Resonance Imaging (fMRI/sMRI). fMRI provides indirect measurements of localised neural activity via blood-oxygenation changes, and sMRI provides detailed anatomy. Joint analysis of EEG with fMRI/sMRI can correlate functional dynamics with structural alterations and improve our understanding of pain mechanisms [23].
- Functional Near-Infrared Spectroscopy (fNIRS). fNIRS measures cortical hemodynamics and provides information complementary to EEG; their joint analysis can shed light on the neurovascular substrate of chronic pain [4].
- Galvanic Skin Response (GSR). GSR reflects sweat-gland activity and indexes emotional and physiological arousal; combined with EEG it can characterise the autonomic response to pain.
- Heart rate (HR) and electrocardiography (ECG). Cardiovascular indices provide information about heart-rate variability and its relation to pain perception.
- Accelerometer (ACC). ACC data capture motion and posture, and are useful in conjunction with EEG for studying motor imagery and motor execution.
- Electromyography (EMG). EMG records muscle electrical activity, and joint EEG–EMG analysis informs the study of cortico-muscular coupling and motor control [17].
- Electrooculography (EOG). EOG records eye movements and is widely used to identify and remove ocular artifacts from EEG; eye-movement features may also reflect cognitive and attentional processes relevant to pain [9].
- Respiratory signals. Breathing patterns are modulated by pain and stress, and their integration with EEG informs the study of respiratory responses to pain.
- Eye-tracking and eye-movement data. Detailed gaze and visual-attention information that, combined with EEG, supports the study of cognitive aspects of pain.
- Positron Emission Tomography (PET). PET provides metabolic and molecular information on brain activity; joint EEG–PET analysis offers a comprehensive view of metabolism and neural dynamics in chronic pain.
- Clinical features. Patient history, physical-examination findings, and treatment-outcome variables enrich EEG-based analyses by providing a holistic view of the patient and improving model accuracy.
- Speech and digitised cognitive parameters. Speech patterns and cognitive assessments provide additional information about mental state and cognitive function; together with EEG they support the study of the cognitive aspects of pain.

Across these modalities, the integration of EEG with EOG, fNIRS, and fMRI is particularly important: each provides complementary information that, when fused, sharpens our understanding of the underlying neural mechanisms and improves the accuracy of predictive models [4, 23].

#### *3.4.2 Relevance of Multimodal Data to the Present Study*

Given that our protocol acquires EEG, EOG, and EMG synchronously during resting-state and motor imagery / motor execution tasks, the integration of these modalities is directly applicable to our work in several ways: (i) enhanced signal interpretation, since combining EEG with imaging modalities such as fMRI and sMRI helps localise activity and link function to structure in chronic pain [23]; (ii) more comprehensive physiological profiling, by including signals such as GSR, HR, and respiration to capture autonomic and respiratory responses; (iii) artifact reduction and signal correction, by using EOG and eye-movement data to clean EEG and improve the validity of downstream analyses [9]; and (iv) multidimensional characterisation of pain, by analysing neural and non-neural signals jointly to capture sensory-discriminative, affective, and cognitive components of pain.

## 4. Discussion

The preprocessing pipeline for both resting-state and motor-imagery / motor-execution EEG has now been implemented end-to-end and validated qualitatively on the available recordings, although it remains under active development as additional data are collected. Our continued investment in this pipeline reflects the methodological care that complex multi-channel EEG data demands.

In parallel, our post-processing strategy has been informed by the structured literature review described in Section 3, which provides a strong empirical basis for model selection and a starting catalogue of candidate architectures. The combination of contemporary published methods and the pipeline-specific constraints of our dataset should yield a robust set of predictive models for analysing EEG across the resting-state and motor conditions.

The next steps include building the full post-processing pipeline (feature extraction, model training, cross-validation, and clinical interpretation), and reporting the resulting methodology and findings in a follow-up technical paper. This continuing effort is aligned with the broader goal of advancing chronic pain management using EEG biomarkers and machine learning.

## 5. Conclusions and Future Work

This work has laid the methodological groundwork for using EEG signals to predict treatment efficacy in chronic neck pain. We implemented and validated a preprocessing pipeline for both resting-state and motor-imagery/motor-execution EEG, which has been reviewed and accepted by our community partner. The pipeline's robustness, evidenced by its capacity to refine raw multi-channel EEG into clean, analysis-ready data, supports its use as the foundation for downstream modelling work.

The ongoing development of post-processing models will exploit the full information content of the preprocessed EEG. Through iterative experimentation and feedback from initial tests, we aim to produce models that are simultaneously accurate, interpretable, and clinically practical.

Our systematic literature-review strategy has been thorough and effective in informing both methodological choices and modelling priorities, and will remain integral to subsequent stages of the project.

Looking ahead, we intend to:

- Develop, train, and validate deep-learning models for treatment-efficacy prediction, prioritising the architectures identified as most promising in our review (e.g., CNN-based and Transformer-based models on spectrotemporal EEG features) [7, 22].

- Maintain ongoing dialogue with our community partner to ensure that models meet the practical needs of patients and clinicians alike.
- Expand the dataset, potentially incorporating multimodal data (EOG, EMG, and where feasible imaging or autonomic signals) to enhance robustness and generalisability.
- Explore a unified model that can ingest all data modalities under a single architecture, as our expertise and computational resources mature.

Through this structured, methodologically explicit approach, we aim to make a meaningful contribution to personalised medicine for chronic pain, advancing both the academic state of the art and the practical needs of the healthcare community.

## Acknowledgements

The authors thank our clinical and community partners for their input on the study design, and the participants who contributed EEG and EMG recordings.

## Author Contributions

X.W., A.N., and Z.G. designed the study. X.W., A.L., H.T., and S.F. carried out data acquisition, preprocessing, and analysis. X.W. drafted the manuscript. A.N. and Z.G. supervised the project. All authors reviewed and approved the final manuscript.

## Conflict of Interest

The authors declare no competing interests.